\documentclass[12pt]{iopart}

\usepackage{graphicx}% Include figure files
\usepackage{bm}% bold math

\begin{document}

\title{Fast unfolding of communities in large networks}

\author{Vincent D. Blondel$^{1;a}$, Jean-Loup Guillaume$^{1,2;b}$, Renaud Lambiotte$^{1,3;c}$ and Etienne Lefebvre$^{1}$}

\address{                    
 $^1$Department of Mathematical Engineering, Universit{\' e} catholique de Louvain, 4 avenue Georges Lemaitre, B-1348 Louvain-la-Neuve, Belgium\\
$^2$ LIP6, Universit{\'e} Pierre et Marie Curie, 4 place Jussieu, 75005 Paris, France\\
 $^3$ Institute for Mathematical Sciences, Imperial College London, 53 Prince's Gate, South Kensington campus, SW72PG, UK
}
\ead{$^a$vincent.blondel@uclouvain.be; $^b$jean-loup.guillaume@lip6.fr; $^c$r.lambiotte@imperial.ac.uk; }

\begin{abstract}
We propose a simple method to extract the community structure of large
networks. Our method is a heuristic method that is based on modularity optimization. It is shown to outperform all other known community detection method in terms of computation time. Moreover, the quality of the
communities detected is very good, as measured by the so-called
modularity. This is shown first by identifying language communities in a
Belgian mobile phone network of 2.6 million customers and by analyzing a web graph of 118 million nodes and more than one billion links. The accuracy of our algorithm is also verified on ad-hoc modular networks. 
\end{abstract}


\maketitle

\section{Introduction}

Social, technological and information systems can often be described in terms of complex networks that have a topology of interconnected nodes combining organization and randomness \cite{AB,NBW}. The typical size of large networks such as social network services, mobile phone networks or the web now counts in millions when not billions of nodes and these scales demand new methods to retrieve comprehensive information from their structure. A promising approach consists in decomposing the networks into sub-units or communities, which are sets of highly inter-connected nodes  \cite{FC}. The identification of these communities is of crucial importance as they may help to uncover a-priori unknown functional modules such as topics in information networks or cyber-communities in social networks. Moreover, the resulting meta-network, whose nodes are the communities, may then be used to visualize the original network structure. 

The problem of community detection requires the partition of a network into communities of densely connected nodes, with the nodes belonging to different communities being only sparsely connected. Precise formulations of this optimization problem are known to be computationally intractable. Several algorithms have therefore been proposed to find reasonably good partitions in a reasonably fast way. This search for fast algorithms has attracted much interest in recent years due to the increasing availability of large network data sets and the impact of networks on every day life. One can distinguish several types of community detection algorithms: divisive algorithms detect inter-community links and remove them from the network \cite{GN,NG,RCC}, agglomerative algorithms merge similar nodes/communities recursively \cite{PL} and optimization methods are based on the maximisation of an objective function \cite{CNM,WH, N06a}. The quality of the partitions resulting from these methods is often measured by the so-called modularity of the partition. The modularity of a partition is a scalar value between -1 and 1 that measures the density of links inside communities as compared to links between communities \cite{GN,N06b}. In the case of weighted networks (weighted networks are networks that have weights on their links, such as the number of communications between two mobile phone users), it is defined as \cite{weighted}
\begin{equation}
Q = {1\over2m} \sum_{i,j} \biggl[ A_{ij} - {k_ik_j\over2m} \biggr]
    \delta(c_i,c_j),
\label{modularity}
\end{equation}
where $A_{ij}$ represents the weight of the edge between $i$ and $j$, $k_i=\sum_j A_{ij}$ is the sum of the weights of the edges attached to vertex $i$,
$c_i$ is the community to which vertex~$i$ is assigned, the $\delta$-function $\delta(u,v)$ is 1 if $u=v$ and 0 otherwise
and $m=\frac{1}{2}\sum_{ij} A_{ij}$.

Modularity has been used to compare the quality of the partitions obtained by different methods, but also as an objective function to optimize \cite{N}. Unfortunately, exact modularity optimization is a problem that is computationally hard \cite{BDG} and so approximation algorithms are necessary when dealing with large networks. The fastest approximation algorithm for optimizing modularity on large networks was proposed by Clauset et al. \cite{CNM}. That method consists in recurrently merging communities that optimize the production of modularity. Unfortunately, this greedy algorithm may produce values of modularity that are significantly lower than what can be found by using, for instance, simulated annealing \cite{guimera}. Moreover, the method proposed in \cite{CNM} has a tendency to produce super-communities that contain a large fraction of the nodes, even on synthetic networks that have no significant community structure. This artefact also has the disadvantage to slow down the algorithm considerably and makes it inapplicable to networks of more than a million nodes. This undesired effect has been circumvented by introducing tricks in order to balance the size of the communities being merged, thereby speeding up the running time and making it possible to deal with networks that have a few million nodes \cite{WT}. 

The largest networks that have been dealt with so far in the literature are a protein-protein interaction network of 30739 nodes \cite{PDF}, a network of about 400000 items on sale on the website of a large on-line retailer \cite{CNM}, and a Japanese social networking systems of about 5.5 million users \cite{WT}. These sizes still leave considerable room for improvement \cite{reka} considering that, as of today, the social networking service Facebook has about 64 million active users, the mobile network operator Vodaphone has about 200 million customers and Google indexes several billion web-pages. Let us also notice that in most large networks such as those listed above there are several natural organization levels --communities divide themselves into sub-communities-- and it is thus desirable to obtain community detection methods that reveal this hierarchical structure \cite{sales}. 

\begin{figure}
\includegraphics[width=1.0\textwidth]{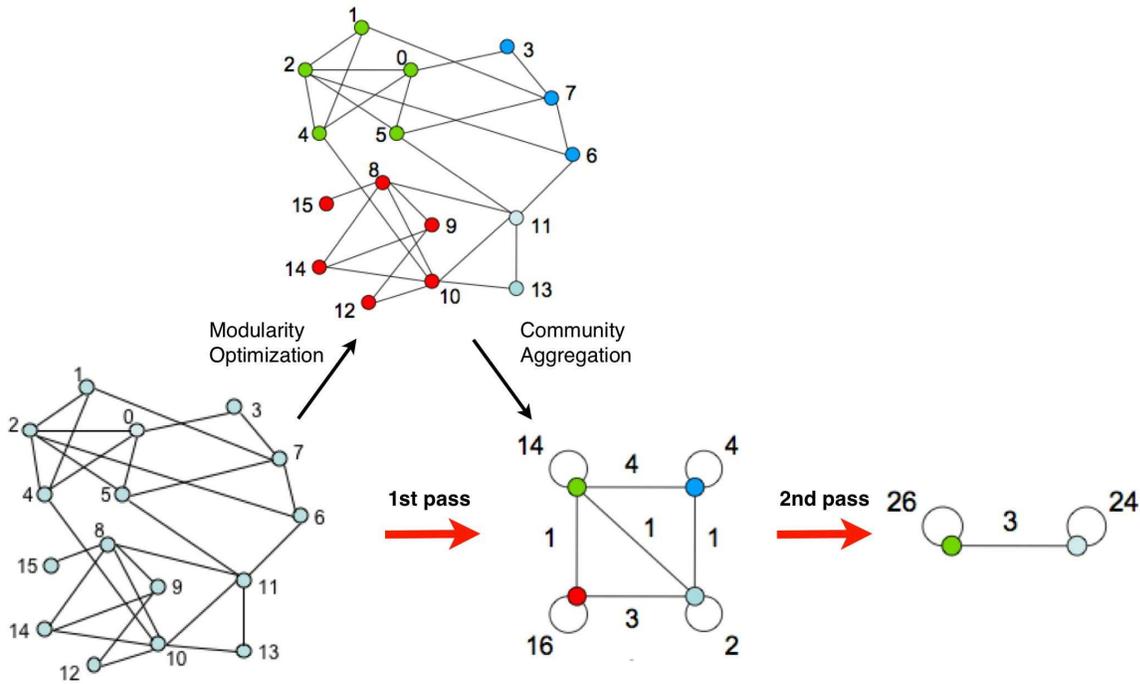}
\caption{Visualization of the steps of our algorithm. Each pass is made of two phases: one where modularity is optimized by allowing only local changes of communities; one where the found communities are aggregated in order to build a new network of communities. The passes are repeated iteratively until no increase of modularity is possible.}\label{fig0}
\end{figure}

\section{Method}
We now introduce our algorithm that finds high modularity partitions of large networks in short time and that unfolds a complete hierarchical community structure for the network, thereby giving access to different resolutions of community detection. Contrary to all the other community detection algorithms, the network size limits that we are facing with our algorithm are due to limited storage capacity rather than limited computation time: identifying communities in a 118 million nodes network took only 152 minutes \cite{note1}. 

Our algorithm is divided in two phases that are repeated iteratively. Assume that we start with a weighted network of $N$ nodes. First, we assign a different community to each node of the network. So, in this initial partition there are as many communities as there are nodes. Then, for each node $i$ we consider the neighbours $j$ of $i$ and we evaluate the gain of modularity that would take place by removing $i$ from its community and by placing it in the community of $j$. The node $i$ is then placed in the community for which this gain is maximum (in case of a tie we use a breaking rule), but only if this gain is positive. If no positive gain is possible, $i$ stays in its original community. This process is applied repeatedly and sequentially for all nodes until no further improvement can be achieved and the first phase is then complete. Let us insist on the fact that a node may be, and often is, considered several times. This first phase stops when a local maxima of the modularity is attained, i.e., when no individual move can improve the modularity. One should also note that the output of the algorithm depends on the order in which the nodes are considered. Preliminary results on several test cases seem to indicate that the ordering of the nodes does not have a significant influence on the modularity that is obtained. However the ordering can influence the computation time. The problem of choosing an order is thus worth studying since it could give good heuristics to enhance the computation time.

Part of the algorithm efficiency results from the fact that the gain in modularity $\Delta Q$ obtained by moving an isolated node $i$ into a community $C$ can easily be computed by:

\begin{eqnarray}
\label{new}
\Delta Q=\left[ \frac{\sum_{in} + k_{i,in}}{2 m} - \left(\frac{\sum_{tot} + k_{i}}{2m}\right)^2 \right] 
- \left[ \frac{\sum_{in} }{2m} - \left(\frac{\sum_{tot}}{2m}\right)^2 - \left(\frac{k_i}{2m}\right)^2 \right],
\end{eqnarray}
where $\sum_{in}$ is the sum of the weights of the links inside $C$,
$\sum_{tot}$  is the sum of the weights of the links incident to nodes in $C$, $k_i$ is the sum of the weights of the links incident to node $i$, $k_{i,in}$ is the sum of the weights of the links from $i$ to nodes in $C$ and $m$ is the sum of the weights of all the links in the network. A similar expression is used in order to evaluate the change of modularity when $i$ is removed from its community. In practice, one therefore evaluates the change of modularity by removing $i$ from its community and then by moving it into a neighbouring community. 

The second phase of the algorithm consists in building a new network whose nodes are now the communities found during the first phase. To do so, the weights of the links between the new nodes are given by the sum of the weight of the links between nodes in the corresponding two communities \cite{arenasf2}. Links between nodes of the same community lead to self-loops for this community in the new network. Once this second phase is completed, it is then possible to reapply the first phase of the algorithm to the resulting weighted network and to iterate. Let us denote by ''pass" a combination of these two phases. By construction, the number of meta-communities decreases at each pass, and as a consequence most of the computing time is used in the first pass. The passes are iterated (see Figure \ref{fig0}) until there are no more changes and a maximum of modularity is attained.
The algorithm is reminiscent of the self-similar nature of complex networks \cite{havlin} and naturally incorporates a notion of hierarchy, as communities of communities are built during the process. The height of the hierarchy that is constructed is determined by the number of passes and is generally a small number, as will be shown on some examples below.

\begin{figure}
\includegraphics[width=1.0\textwidth]{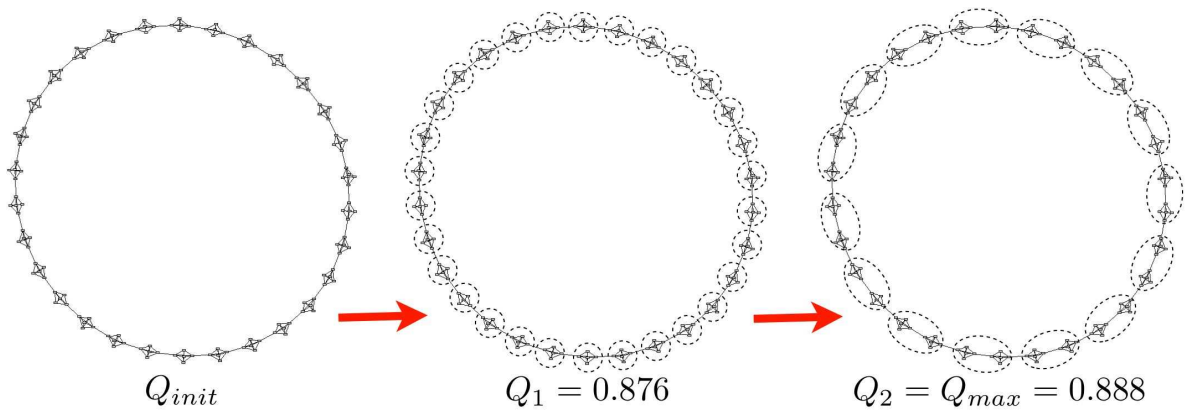}
\caption{We have applied our method to the ring of 30 cliques discussed in \cite{FB}. The cliques are composed of 5 nodes and are inter-connected through single links. The first pass of the algorithm finds the natural partition of the network. The second pass finds the global maximum of modularity where cliques are combined into groups of two. }\label{resolution}
\end{figure}

\begin{figure}
\includegraphics[width=0.8\textwidth]{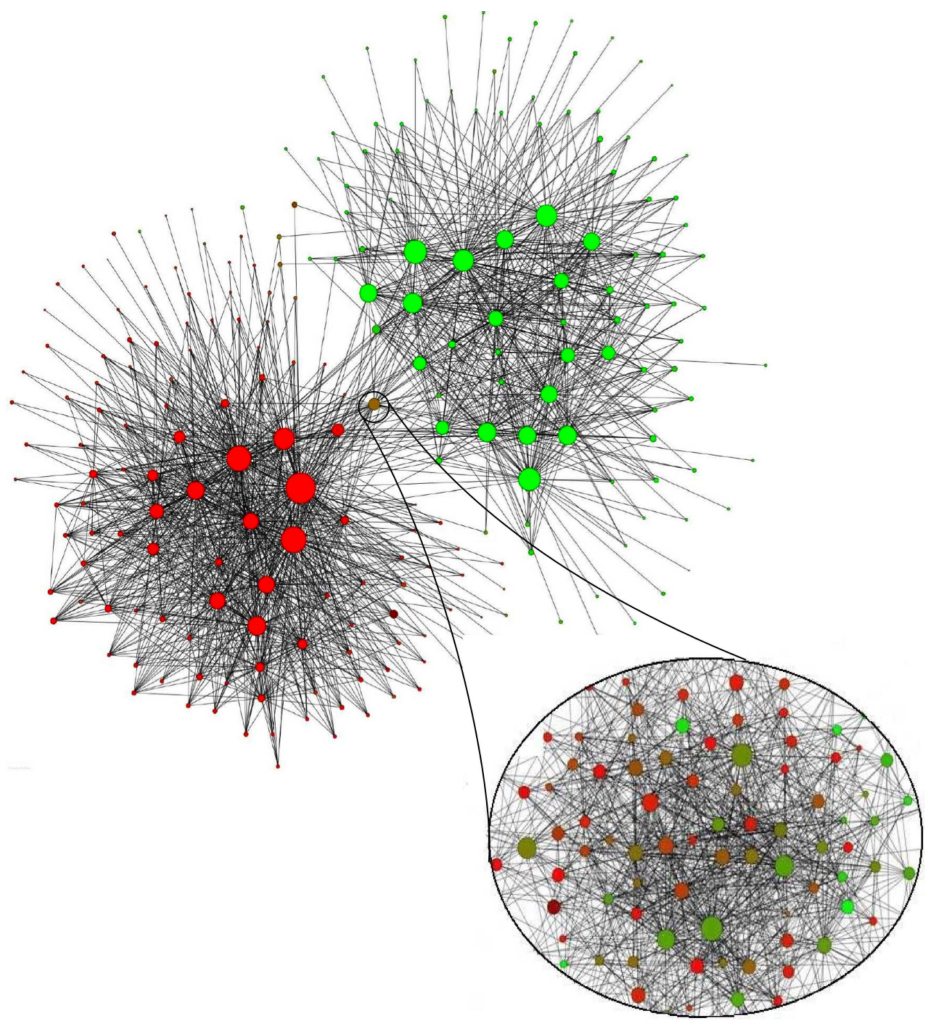}
\caption{Graphical representation of the network of communities extracted from a Belgian mobile phone network. About 2M customers are represented on this network. The size of a node is proportional to the number of individuals in the corresponding community and its colour on a red-green scale represents the main language spoken in the community (red for French and green for Dutch). Only the communities composed of more than 100 customers have been plotted. Notice the intermediate community of mixed colours between the two main language clusters. A zoom at higher resolution reveals that it is made of several sub-communities with less apparent language separation.}\label{fig3}
\end{figure}

\begin{figure}
\includegraphics[width=0.8\textwidth]{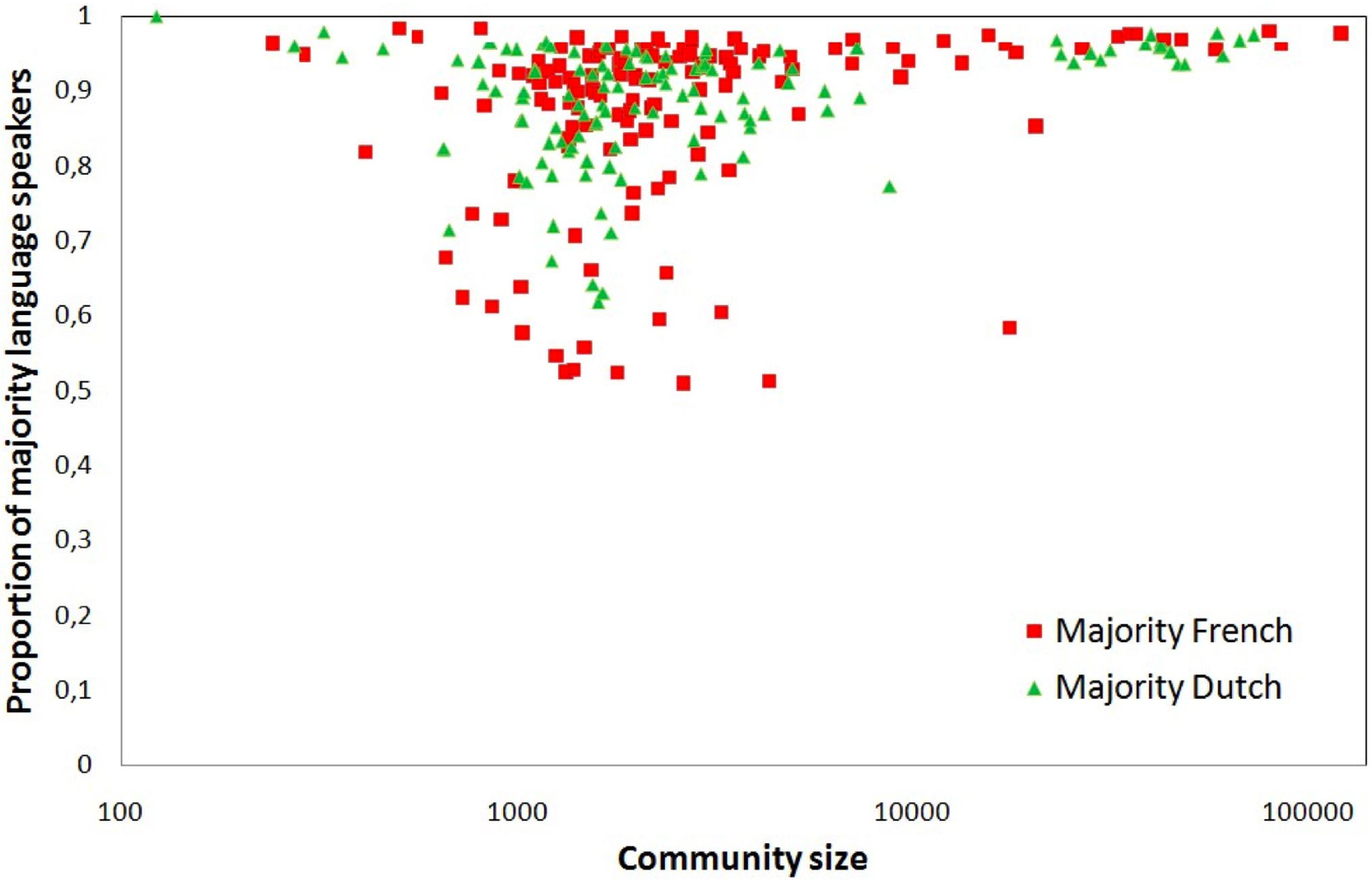}
\caption{For the largest communities in the Belgian mobile phone network we represent the size of the community and the proportion of customers in the community that speak the dominant language of the community. For all but one community of more than 10000 members the dominant language is spoken by more than 85\% of the community members.}\label{fig4}
\end{figure}

This simple algorithm has several advantages. First, its steps are intuitive and easy to implement, and the outcome is unsupervised. Moreover, the algorithm is extremely fast, i.e., computer simulations on large ad-hoc modular networks suggest that its complexity is linear on typical and sparse data. This is due to the fact that the possible gains in modularity are easy to compute with the above formula and that the number of communities decreases drastically after just a few passes so that most of the running time is concentrated on the first iterations. The so-called resolution limit problem of modularity also seems to be circumvented thanks to the intrinsic multi-level nature of our algorithm. Indeed, it is well-known  \cite{FB} that modularity optimization fails to identify communities smaller than a certain scale, thereby inducing a resolution limit on the community detected by a pure modularity optimization approach. This observation is only partially relevant in our case because the first phase of our algorithm involves the displacement of single nodes from one community to another. Consequently, the probability that two distinct communities can be merged by moving nodes one by one is very low. These communities may possibly be merged in the later passes, after blocks of nodes have been aggregated. However, our algorithm provides a decomposition of the network into communities for different levels of organization. For instance, when applied on the clique network proposed in \cite{FB}, the cliques are indeed merged in the final partition but they are distinct after the first pass (see Figure \ref{resolution}). This result suggests that the intermediate solutions found by our algorithm may also be meaningful and that the uncovered hierarchical structure may allow the end-user to zoom in the network and to observe its structure with the desired resolution. 

\section{Application to large networks}
\label{application}

\begin{table*}[b]
\begin{tabular*}{\textwidth}{@{\extracolsep{\fill}}rrrrrrrr}
\hline
\multicolumn1c{}&\multicolumn1c{\footnotesize Karate}&
\multicolumn1c{\footnotesize Arxiv}&\multicolumn1c{\small Internet}&
\multicolumn1c{\footnotesize Web nd.edu}&\multicolumn1c{\footnotesize Phone}&
\multicolumn1c{\footnotesize Web uk-2005}&\multicolumn1c{\footnotesize Web WebBase 2001}\cr
\hline
\multicolumn1c{\footnotesize Nodes/links}&\multicolumn1c{\footnotesize 34/77}&
\multicolumn1c{\footnotesize 9k/24k}&\multicolumn1c{\footnotesize 70k/351k}&
\multicolumn1c{\footnotesize 325k/1M}&\multicolumn1c{\footnotesize 2.6M/6.3M}&
\multicolumn1c{\footnotesize 39M/783M}&\multicolumn1c{\footnotesize 118M/1B}\cr
\hline
\multicolumn1c{\footnotesize CNM}&\multicolumn1c{\footnotesize .38/0s}&
\multicolumn1c{\footnotesize .772/3.6s}&\multicolumn1c{\footnotesize .692/799s}&
\multicolumn1c{\footnotesize .927/5034s}&\multicolumn1c{-/-}&
\multicolumn1c{-/-}&\multicolumn1c{-/-}\cr
\hline
\multicolumn1c{\footnotesize PL}&\multicolumn1c{\footnotesize .42/0s}&
\multicolumn1c{\footnotesize .757/3.3s}&\multicolumn1c{\footnotesize .729/575s}&
\multicolumn1c{\footnotesize .895/6666s}&\multicolumn1c{-/-}&
\multicolumn1c{-/-}&\multicolumn1c{-/-}\cr
\hline
\multicolumn1c{\footnotesize WT}&\multicolumn1c{\footnotesize .42/0s}&
\multicolumn1c{\footnotesize .761/0.7s}&\multicolumn1c{\footnotesize .667/62s}&
\multicolumn1c{\footnotesize .898/248s}&\multicolumn1c{\footnotesize .56/464s}&
\multicolumn1c{-/-}&\multicolumn1c{-/-}\cr
\hline
\multicolumn1c{\footnotesize Our algorithm}&\multicolumn1c{\footnotesize .42/0s}&
\multicolumn1c{\footnotesize .813/0s}&\multicolumn1c{\footnotesize .781/1s}&
\multicolumn1c{\footnotesize .935/3s}&\multicolumn1c{\footnotesize .769/134s}&
\multicolumn1c{\footnotesize .979/738s}&\multicolumn1c{\footnotesize .984/152mn}\cr
\hline
\end{tabular*}
\caption{Summary of numerical results. This table gives the performances of the algorithm of Clauset, Newman and Moore \cite{CNM}, of Pons and Latapy \cite{PL}, of Wakita and Tsurumi \cite{WT} and of our algorithm for community detection in networks of various sizes. For each method/network, the table displays the modularity that is achieved and the computation time. Empty cells correspond to a computation time over 24 hours. Our method clearly performs better in terms of computer time and modularity. It is also interesting to note the small value of $Q$ found by WT for the mobile phone network. This bad modularity result may originate from their heuristic which creates balanced communities, while our approach gives unbalanced communities in this specific network.}
\end{table*}

In order to verify the validity of our algorithm, we have applied it on a number of test-case networks that are commonly used for efficiency comparison  and we have compared it with three other community detection algorithms (see Table 1). The networks that we consider include a small social network \cite{Z}, a network of 9000 scientific paper and their citations \cite{1}, a sub-network of the internet \cite{HM} and a webpage network of a few hundred thousands web-pages (the Ònd.eduÓ domain, see \cite{AJB}). In all cases, one can observe both the rapidity and the large values of the modularity that are obtained. Our method outperforms all the other methods to which it is compared. We also have applied our method on two web networks of unprecedented sizes: a sub-network of the Ò.ukÓ domain of 39 million nodes and 783 million links \cite{2} and a network of 118 million nodes and 1 billion links obtained by the Stanford WebBase crawler \cite{2,3}. Even for these very large networks, the computation time is small (12 minutes and 152 minutes respectively) and makes networks of still larger size, perhaps a billion nodes, accessible to computational analysis. It is also interesting to note that the number of passes is usually very small. In the case of the Karate Club  \cite{Z}, for instance, there are only 3 passes: during the first one, the 34 nodes of the network are partitioned into 6 communities; after the second one, only four communities remain; during the third one, nothing happens and the algorithm therefore stops. In the above examples, the number of passes is always smaller than 5.

We have also tested the sensitivity of our algorithm by applying it on ad-hoc networks that have a known community structure. To do so, we have used networks composed of $128$ nodes which are split into 4 communities of 32 nodes each \cite{danon}. Pairs of nodes belonging to the same community are linked with probability $p_{in}$ while pairs belonging to different communities are linked with probability $p_{out}$. The accuracy of the method is evaluated by measuring the fraction of correctly identified nodes and the normalized mutual information. In the benchmark proposed in \cite{danon}, the fraction of correctly identified nodes is $0.67$ for $z_{out}=8$, $0.92$ for $z_{out}=7$ and $0.98$ for $z_{out}=6$, i.e., an accuracy similar to that of the algorithm of Pons and Latapy \cite{PL} and of the algorithm of Reichardt and Bornholdt \cite{RB}. To our knowledge, only two algorithms have a better accuracy than ours, the algorithm of Duch and Arenas \cite{DA} and the simulated annealing method first proposed in \cite{guimera}, but their computational cost limits their applicability to much smaller networks than the ones considered here. Our algorithm has also been successfully tested on other benchmarks, such as the ones proposed in \cite{sales,fortunato}. In the benchmark proposed in \cite{fortunato}, for instance, the normalized mutual information is nearly $1$ for the macro-communities with a mixing parameter $k_3$ up to $35$. It reaches $0.5$ when the mixing parameter is around $55$.

To validate the communities obtained we have also applied our algorithm to a large network constructed from the records of a Belgian mobile phone company. This network is described in details in \cite{LDB} where it is shown to exhibits typical features of social networks, such as a high clustering coefficient and a fat-tailed degree distribution. The network is composed of 2.6 million customers, between whom weighted links are drawn that account for their total number of phone calls during a 6 month period. Each customer is identified by a surrogate key to which several entries are associated, such as his age, his sex, his language and the zip code of the place where he lives. This large social network is exceptional due to the particular situation of Belgium where two main linguistic communities (French and Dutch) coexist and which provides an excellent way to test the validity of our community detection method by looking at the linguistic homogeneity of communities \cite{PBV}. From a more sociological point of view, the possibility to highlight the linguistic, religious or ethnic homogeneity of communities opens perspectives for describing the social cohesion and the potential fragility of a country \cite{onnela}. 

On this particular network, our community detection algorithm has identified a hierarchy of six levels. At the bottom level every customer is a community of its own and at the top-level there are 261 communities that have more than 100 customers. These communities account for about 75\% of all customers. We have performed a language analysis of these 261 communities (see Figure \ref{fig3}). The homogeneity of a community is characterized by the percentage of those speaking the dominant language in that community; this quantity goes to 1 when the community tends to be monolingual. Our analysis reveals that the network is strongly segregated, with most communities almost monolingual.  There are 36 communities with more than 10000 customers and, except for one community at the interface between the two language clusters, all these communities have more than 85\% of their members speaking the same language (see Figure \ref{fig4} for a complete distribution). It is interesting to analyse more closely the only community that has a more equilibrate distribution of languages. Our hierarchy revealing algorithm allows us to do this by considering the sub-communities provided by the algorithm at the lower level. As shown on Figure 4, these sub-communities are closely connected to each other and are themselves composed of heterogeneous groups of people. These groups of people, where language ceases to be a discriminating factor, might possibly play a crucial role for the integration of the country and for the emergence of consensus between the communities \cite{lambi}. One may indeed wonder what would happen if the community at the interface between the two language clusters on Figure 3 was to be removed. 

Another interesting observation is related to the presence of other languages. There are actually four possible language declarations for the customers of this particular mobile phone operator: French, Dutch, English or German. It is interesting to note that, whereas English speaking customers disperse themselves quite evenly in all communities, more than 60\% of the German speaking customers are concentrated in just one community. This is probably due to the fact that German speaking people are mainly concentrated in a small region close to Germany, while English speaking people are spread in the whole country.
Let us finally observe that, as can be visually noticed on Figure ~3, French speaking communities are much more densely connected than their Dutch speaking counterparts: on average, the strength of the links between French speaking communities is 54\% stronger than those between Dutch speaking communities. This difference of structure between the two sub-networks seems to indicate that the two linguistic communities are characterized by different social behaviours and therefore suggests to search other topological characteristics for the communities. 

\section{Conclusion and discussion}

We have introduced an algorithm for optimizing modularity that allows to study networks of unprecedented size. The limitation of the method for the experiments that we performed was the storage of the network in main memory rather than the computation time.  This change of scales, i.e., from around 5 millions nodes for previous methods to more than 100 millions nodes in our case, opens exciting perspectives as the modular structure of complex systems such as whole countries or huge parts of the Internet can now be unraveled. The accuracy of our method has also been tested on ad-hoc modular networks and is shown to be excellent in comparison with other (much slower) community detection methods. It is interesting to note that the speed of our algorithm can still be substantially improved by using some simple heuristics, for instance by stopping the first phase of our algorithm when the gain of modularity is below a given threshold or by removing the nodes of degree 1 (leaves) from the original network and adding them back after the community computation. The impact of these heuristics on the final partition of the network should be studied further, as well as the role played by the ordering of the nodes during the first phase of the algorithm. 

By construction, our algorithm unfolds a complete hierarchical community structure for the network, each level of the hierarchy being given by the intermediate partitions found at each pass. In this paper, however, we have only verified the accuracy of the top level of this hierarchy, namely the final partition found by our algorithm, and the accuracy of the intermediate partitions has still to be shown. Several points suggest, however, that these intermediate partitions make sense. First, intermediate partitions correspond to local maxima of modularity, maxima in the sense that it is not possible to increase modularity by moving one single "entity" from one community to a neighbouring one. In the first pass of the algorithm, these entities are nodes, but at subsequent passes, they correspond to larger and larger sets of nodes. Intermediate partitions may therefore be viewed as local maxima of modularity at different scales. It is the agglomeration of nodes during the second phase of the algorithm which allows to uncover larger and larger communities, thereby taking advantage of the self-similar structure of many complex networks. Second, the final partition found by our algorithm has a very high value of modularity for a broad range of system sizes (for instance, as shown in Table 1, our algorithm performs better in terms of modularity than those of Clauset, Newman and Moore \cite{CNM}, of Pons and Latapy \cite{PL} and of Wakita and Tsurumi \cite{WT}).  Finally, it is instructive to consider a community $C$ found at the last pass of our algorithm. In order to test the validity of the sub-communities found at the penultimate pass, it is tempting to look at community $C$ as a new network, thereby neglecting links going from $C$ to the rest of the network. By reapplying our algorithm on the isolated community $C$, one expects to find very similar sub-communities due to the local optimization involved at each step. These are, however, very qualitative arguments and the multi-resolution of our algorithm will only be confirmed after looking in detail at the hierarchies found in ad-hoc networks with known hierarchical structure \cite{sales} or without community structure (e.g. Erd{\"o}s-Renyi random graphs), or after comparing with other methods incorporating a tunable resolution \cite{fortunato,arenasf,delvenne}. 

 {\bf Acknowledgements}
This research was supported by the Communaut{\'e} Fran\c{c}aise de Belgique through a grant ARC and by the Belgian Network DYSCO, funded by the Interuniversity Attraction Poles Programme, initiated by the Belgian State, Science Policy Office.  J.-L. G. is also supported by the project MAPE (ANR France) and MAPAP (Safer Internet Plus Programme, European Union).

\end{document}